\documentclass[twocolumn,pra,groupedaddress,superscriptaddress,nofootinbib]{revtex4}
\usepackage{graphicx,amsmath,amssymb,txfonts,float}
\usepackage{subfigure,hyperref,bbm,times}
\usepackage[T1]{fontenc}
\usepackage{braket}
\usepackage{epsfig}
\usepackage{color}
\usepackage{graphicx}
\usepackage{dcolumn}
\usepackage{bm}
\usepackage{soul}

\usepackage[usenames]{xcolor}

\renewcommand{\tilde}{~}

\newcommand{\1}{\mathds{1}}

\newcommand{\phis}{\phi_s}

\begin{document}
	
	\title{Direct measurement of particle statistical phase}
	
	\author{Yan Wang}
	\affiliation{CAS Key Laboratory of Quantum Information, University of Science and Technology of China, Hefei 230026, People's Republic of China}
	\affiliation{Synergetic Innovation Center of Quantum Information and Quantum Physics, University of Science and Technology of China, Hefei 230026, People's Republic of China}
	
	\author{Matteo Piccolini}
	\email{matteo.piccolini@unipa.it}
	\affiliation{Dipartimento di Ingegneria, Universit\`{a} di Palermo, Viale delle Scienze, 90128 Palermo, Italy}
	\affiliation{INRS-EMT, 1650 Boulevard Lionel-Boulet, Varennes, Qu\'{e}bec J3X 1S2, Canada}
	
	\author{Ze-Yan Hao}
	\affiliation{CAS Key Laboratory of Quantum Information, University of Science and Technology of China, Hefei 230026, People's Republic of China}
	\affiliation{Synergetic Innovation Center of Quantum Information and Quantum Physics, University of Science and Technology of China, Hefei 230026, People's Republic of China}
	
	\author{Zheng-Hao Liu}
	\affiliation{CAS Key Laboratory of Quantum Information, University of Science and Technology of China, Hefei 230026, People's Republic of China}
	\affiliation{Synergetic Innovation Center of Quantum Information and Quantum Physics, University of Science and Technology of China, Hefei 230026, People's Republic of China}
	
	\author{Kai Sun}
	\email{ksun678@ustc.edu.cn}
	\affiliation{CAS Key Laboratory of Quantum Information, University of Science and Technology of China, Hefei 230026, People's Republic of China}
	\affiliation{Synergetic Innovation Center of Quantum Information and Quantum Physics, University of Science and Technology of China, Hefei 230026, People's Republic of China}
	
	\author{Jin-Shi Xu}
	\email{jsxu@ustc.edu.cn}
	\affiliation{CAS Key Laboratory of Quantum Information, University of Science and Technology of China, Hefei 230026, People's Republic of China}
	\affiliation{Synergetic Innovation Center of Quantum Information and Quantum Physics, University of Science and Technology of China, Hefei 230026, People's Republic of China}
	
	\author{Chuan-Feng Li}
	\email{cfli@ustc.edu.cn}
	\affiliation{CAS Key Laboratory of Quantum Information, University of Science and Technology of China, Hefei 230026, People's Republic of China}
	\affiliation{Synergetic Innovation Center of Quantum Information and Quantum Physics, University of Science and Technology of China, Hefei 230026, People's Republic of China}
	
	\author{Guang-Can Guo}
	\affiliation{CAS Key Laboratory of Quantum Information, University of Science and Technology of China, Hefei 230026, People's Republic of China}
	\affiliation{Synergetic Innovation Center of Quantum Information and Quantum Physics, University of Science and Technology of China, Hefei 230026, People's Republic of China}
	
	\author{Roberto Morandotti}
	\affiliation{INRS-EMT, 1650 Boulevard Lionel-Boulet, Varennes, Qu\'{e}bec J3X 1S2, Canada}
	
	\author{Giuseppe Compagno}
	\affiliation{Dipartimento di Fisica e Chimica - Emilio Segr\`e, Universit\`a di Palermo, via Archirafi 36, 90123 Palermo, Italy}
	
	\author{Rosario Lo Franco}
	\email{rosario.lofranco@unipa.it}
	\affiliation{Dipartimento di Ingegneria, Universit\`{a} di Palermo, Viale delle Scienze, 90128 Palermo, Italy}

\begin{abstract}
The symmetrization postulate in quantum mechanics is formally reflected in the appearance of an exchange phase ruling the symmetry of identical particle global states under particle swapping.
    Many indirect measurements of this fundamental phase have been reported so far, while a direct observation has been only recently achieved for photons.
    Here we propose a general scheme capable to directly measure the exchange phase of any type of particles (bosons, fermions, anyons), exploiting the operational framework of spatially localized operations and classical communication.
    We experimentally implement it in an all-optical platform providing proof-of-principle for different simulated exchange phases. As a byproduct, we supply a direct measurement of the real bosonic exchange phase of photons.
    Also, we analyze the performance of the proposed scheme when mixtures of particles of different nature are injected. Our results confirm the symmetrization tenet and provide a tool to explore it in various scenarios.
    \end{abstract}

	\date{\today }

	\maketitle

The symmetrization postulate divides particles living in a 3-dimensional space into two groups: bosons and fermions. Such postulate forces the state of an ensemble of identical bosons (fermions) to be symmetric (antisymmetric) under the exchange of any pair of particles \cite{Peresbook}. Considering a system of two identical particles, its global state must then satisfy $\ket{\psi(1,2)}=e^{i\phi}\ket{\psi(2,1)}$, where $1$ and $2$ refer to the two constituents and the relative phase $\phi$ is the particle exchange phase (EP), being $\phi=0$ for bosons and $\phi=\pi$ for fermions. Furthermore, the existence of particles called anyons living in 2-dimensional spaces with a fractional EP $\phi\in(0,2\pi)\setminus(\pi)$ has been suggested \cite{leinaas1977theory,wilczek1982magnetic}, attracting the attention of the scientific community in the last decades\tilde\cite{nayak2008non,nakamura2020direct,bartolomei2020fractional}.
Despite the fundamental importance of the symmetrization postulate in both understanding the quantum world and practical applications, its observation has been typically limited to indirect measurements\tilde\cite{Hilborn, Modugno1998, English2010,deilamian_1995, DeAngelis1996, Ospelkaus2010}; only the bosonic nature of photons has been so far directly proven by a state transport protocol \cite{tschernig2021direct,lo2021directly}. A direct observation of fermionic and anyonic EPs is still missing, leaving the field open to the introduction of new techniques capable to fill that gap.
	
In the standard approach to identical particles\tilde\cite{Peresbook} the global state vector is symmetrized/antisymmetrized with respect to unphysical labels associated to each constituent.
This approach is known to exhibit drawbacks when trying to asses real quantum correlations between constituents \cite{tichy,ghirardi}. Given the key role played by entanglement in quantum technologies, different methods have been developed to fix such an issue\tilde\cite{ghirardi,balachandranPRL,sasaki2011PRA,benatti2012bipartite,chin2019entanglement,nolabelappr,compagno2018dealing,slocc}. Among these, the no-label approach \cite{nolabelappr} provides some advantages: it straightforwardly identifies physical entanglement and establishes its quantitative relation with the degree of spatial indistinguishability \cite{indistentanglprotection}; the latter is associated to the spatial overlap of particle wave functions. Importantly, in the no-label formalism, the role played by the particles' nature does not manifest itself in the (anti)symmetrization of the quantum state but in the probability amplitudes of the global system \cite{nolabelappr,compagno2018dealing}.
	
\begin{figure*}[t]
		\centering
		\includegraphics[width=0.98\textwidth]{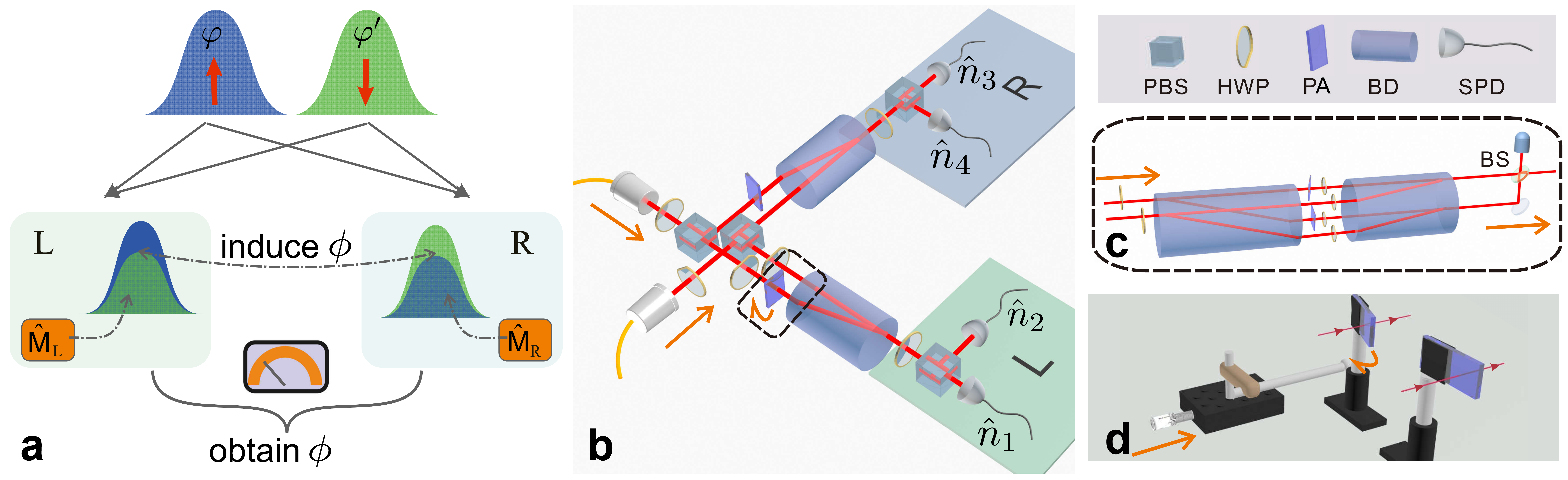}
		\caption{\textbf{Theoretical scheme and experimental setup.} 
{\bf a.} Conceptual procedure. The wave functions of two identical particles are distributed over two distinct regions L and R and made to spatially overlap, generating spatial indistinguishability. A sLOCC measurement (detection + postselection) is performed by coincidence counting between the two regions, to directly measure the EP using a single-particle rotation $\hat{M}$ in the two regions.
{\bf b.} Experimental setup.
Two independently prepared photons with opposite polarizations follow two distinct paths where they pass through an half-wave plate (HWP, one fixed at $22.5^\circ$ and the other at $-\beta/2$), a polarization beam splitter (PBS) directing the beams over two distinct spatial regions L and R. A pair of beam displacers (BDs), one set on L and one on R, merge the two beams over the corresponding region, generating spatial indistinguishability between the two photons.
The relative phase between the two spatial modes of one photon is properly tuned using a phase adjustment (PA) shown in {\bf d} to simulate the desired particle nature.
The four outputs are individually directed towards four single-photon detectors (SPDs), where a proper coincidence measurement is implemented to extrapolate the EP.
{\bf c.} Replacement setup for the dashed frame in {\bf b} used to prepare the mixed states. Another BD divides the corresponding beam in two paths with a $3mm$ vertical separation. The two upper arms are used to simulate the particles with EP $\phi_1$, while the two lower arms prepare the particles with simulated EP $\phi_2$. A subsequent BD merges the two upper arms and the two lower ones into two beams which are finally combined using one beam splitter (BS) with one blocked output.}
	\label{expe}
	\end{figure*}

The no-label approach has been widely exploited within the spatially localized operations and classical communication (sLOCC) environment\tilde\cite{slocc,Castellini2019metrology,indistentanglprotection,Piccolini_2021,indistdynamicalprotection,experimentalslocc,Barros:20,Sun_PhaseDiscr}. Such procedure can be seen as a natural extension of standard local operations and classical communication (LOCC) for distinguishable particles to the case of indistinguishable and individually unaddressable constituents. Operationally, sLOCC make the global state of indistinguishable particles undergo a projective measurement over spatially-separated regions, followed by a post-selection when one particle is found in each location.
Consider a state of two independent identical qubits $\ket{\psi_\textrm{D}}=\ket{\varphi_\textrm{D}\uparrow,\varphi^\prime_\textrm{D}\downarrow}$, where $\varphi_\textrm{D}$, $\varphi'_\textrm{D}$ are spatial wave functions and $\uparrow,\downarrow$ are pseudospins. The result of sLOCC onto $\ket{\psi_\textrm{D}}$ gives \cite{slocc}
    \begin{equation}
	\label{psilr}
		\ket{\psi_{\rm LR}}
		=\frac{lr^\prime\ket{\rm L\uparrow,\rm R\downarrow}+e^{i\phi}rl^\prime\ket{\rm L\downarrow,\rm R\uparrow}}
		{\sqrt{|lr^\prime|^2+|rl^\prime|^2}},
	\end{equation}
where $l,l'$ ($r,r'$) are the probability amplitudes for each particle to be found in the region L (R), while $\phi$ is the exchange (statistical) phase; $\ket{\psi_{\rm LR}}$ is entangled only if the qubits spatially overlap, i.e., are spatially indistinguishable, at the regions L, R (see Appendix \ref{app:Theory} for details). Remarkably, the sLOCC process makes particle statistics naturally emerge in the final entangled state. The entanglement so obtained is experimentally accessible \cite{experimentalslocc,Barros:20}, and has been exploited for teleportation \cite{experimentalslocc} and phase discrimination \cite{Sun_PhaseDiscr}. Also, sLOCC-based indistinguishability is useful for protecting entanglement against noise \cite{indistentanglprotection,Piccolini_2021,indistdynamicalprotection}.
	
Here we give further value to sLOCC by experimentally showing, in a quantum-optical setup, that its theoretical framework enables a phase-estimation procedure to directly access the EP of indistinguishable particles of any nature (Fig.~\ref{expe}\textbf{a}). 

Denoting with $\ket{H}$ and $\ket{V}$ the horizontal and vertical polarization of a photon, respectively, we make the correspondence $\ket{H}\leftrightarrow\ket{\uparrow}$ and $\ket{V}\leftrightarrow\ket{\downarrow}$. A pulsed ultraviolet beam with wavelength at 400 $nm$ is used to pump a type-II phase-matched $\beta$-barium borate (BBO) crystal to generate two uncorrelated photons ($\ket{H}\otimes\ket{V}$) via spontaneous parametric down conversion. Hong-Ou-Mandel interference is performed to characterize the indistinguishability of the two photons, providing a visibility of 97.7\% \cite{experimentalslocc}. Single-mode fibers collect the photons via fiber couplers and direct them towards the effective experimental setup illustrated in 
Fig.~\ref{expe}{\bf b}. Here, the weights of their horizontal and vertical polarizations are tuned using a pair of half wave plates (HWPs) fixed at $22.5^\circ$ and $-\beta/2$ (adjustable angle), respectively. Two polarization beam splitters (PBSs) separate the polarization components of the two photons and direct them towards two distinct regions L and R. An additional pair of HWPs at $45^\circ$ is placed on L to restore the original input polarizations. The result is the preparation of the state $\ket{\psi_\textrm{D}}$ with $\ket{\varphi_D}=(\ket{\rm L}+\ket{\rm R})/\sqrt{2}$,  $\ket{\varphi^\prime_D}=\sin\beta\ket{\rm L}+\cos\beta\ket{\rm R}$.
	
\begin{figure}[t]
		\centering
		\includegraphics[width=0.48\textwidth]{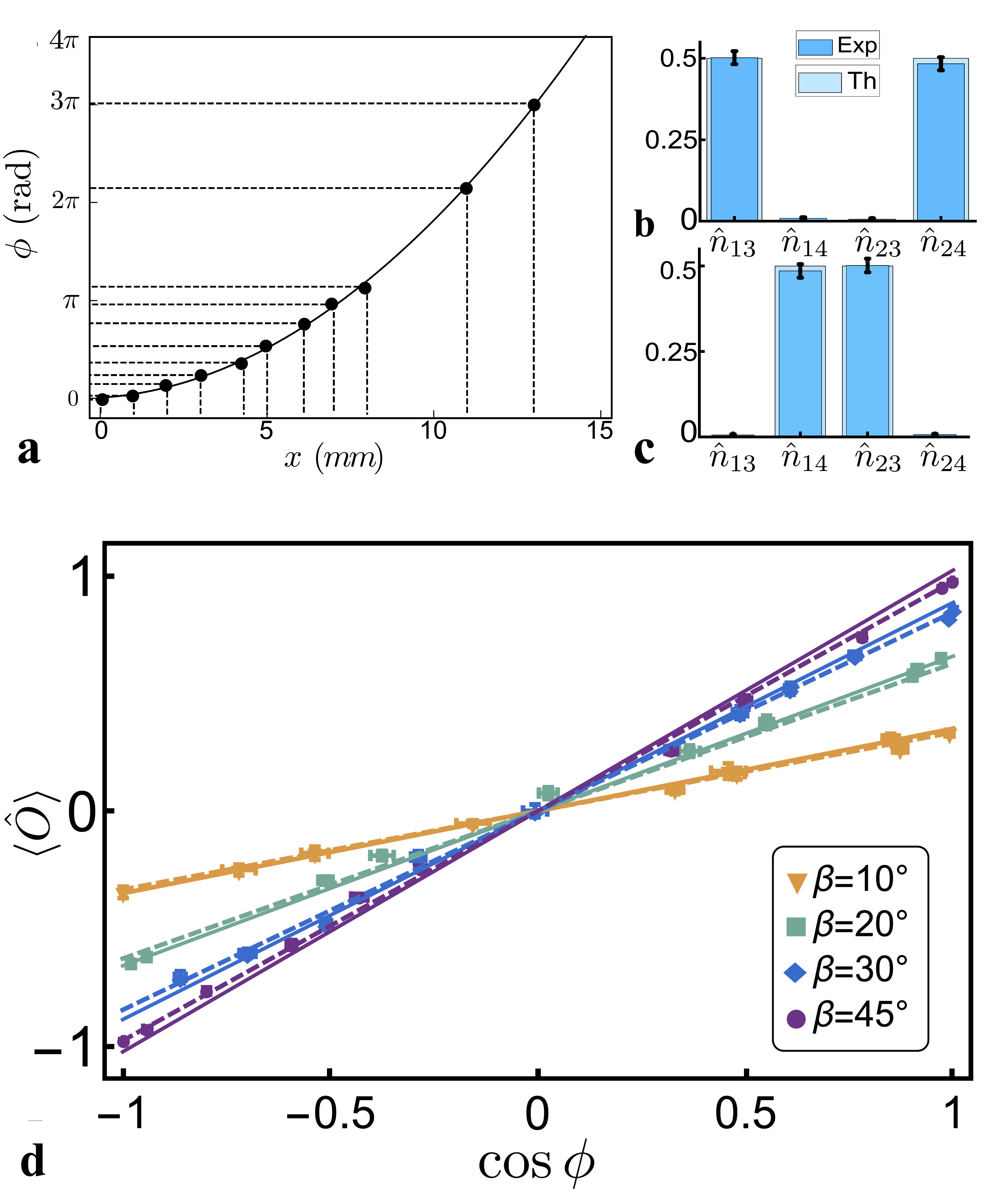}
		\caption{{\bf a.} Relation between the distance $x$ of the moving plate and the generated relative phase $\phi$. The black dots represent the experimental results, while the black line is the theoretical prediction. Errorbars are too small to be visible. Panels {\bf b} and {\bf c} represent the coincidence counting $\hat{n}_{13}$, $\hat{n}_{14}$, $\hat{n}_{23}$, and $\hat{n}_{24}$ measured for bosons and fermions, respectively. {\bf d.} Relation between $\langle \hat{O}\rangle$ and $\cos\phi$, with $\phi$ obtained via quantum tomography.
		Results are reported for different values of $\beta$ (different colors), corresponding to different degrees of spatial indistinguishability. The solid lines represent the theoretically expected results in the ideal (no noise) scenario, while the dashed lines show the theoretical values when noise is taken into consideration. Experimentally measured values are represented with markers.}\label{fig3}
	\end{figure}

Using a home-made phase adjustment (PA) composed of a thin plate of fused quartz fixed in R and of another identical plate tilted and placed in L (Fig.~\ref{expe}{\bf b}, {\bf d}), an arbitrary relative phase $\phis$ is judiciously introduced between the components L and R of the photon $\varphi^\prime_D$, which becomes $\ket{\varphi^\prime_D}=e^{i\phis}\sin\beta\ket{\rm L}+\cos\beta\ket{\rm R}$. As shown in Fig.~\ref{expe}{\bf d}, $\phis$ is tuned by directly adjusting the distance $x$ ($mm$) of a movable plate. The relation between $\phis$ and $x$ is displayed in Fig.\tilde\ref{fig3}{\bf a} with experimental results (dots) and theoretical prediction (solid line) for a plate's thickness $d=199.94\pm1.43\ \mu m$ and a rotation radius $r=102.36\pm0.91\ mm$ (Appendix \ref{app:PA} for more details).

A beam displacer (BD) is used to make the two beams overlap over both regions. We proceed by setting an HWP at $22.5^\circ$ after the BD in both L and R to implement the desired rotation, producing the final state $\ket{\psi_f}$ (Appendix \ref{app:Theory} for details). The pseudospin measurement $\hat{\sigma}_z^{(L)}\otimes\hat{\sigma}_z^{(R)}$ is then performed as a coincidence counting by placing a PBS on both L and R, whose outputs are individually directed towards two different SPDs. The corresponding measured observable is
    \begin{equation}
        \hat{O}=\hat{n}_{13}+\hat{n}_{24}-\hat{n}_{14}-\hat{n}_{23},
    \end{equation}
where $\hat{n}_{ij}=\hat{n}_{i}\hat{n}_{j}$ is the coincidence between the outputs $i$ and $j$,  $\hat{n}_j$ being the number operators related to the four outputs $j=1,\dots,4$ numbered as in Fig.~\ref{expe}{\bf b}.
This spatially localized operation (sLO), made through local counting in L and R, and classical communication (CC) tools, realized via coincidence counts, create the state of Eq.\tilde\eqref{psilr} with $l=r=1/\sqrt{2}$, $l^\prime=\sin\beta$, $r^{\prime}=\cos\beta$, i.e.
    \begin{equation}
    \label{exppsilr}
        \ket{\psi_{\rm LR}}=\cos{\beta}\ket{{\rm L} H,{\rm R} V}+e^{i\phis}\sin{\beta}\ket{{\rm L} V,{\rm R} H},
    \end{equation}
    before the final rotation transforms it into $\ket{\psi_f}$.
Notice that the relative phase $\phis$ in Eq.\tilde\eqref{exppsilr} plays the exact same role of the real EP $\phi$ in Eq.\tilde\eqref{psilr} (which here is set to zero since our experiment is run by using bosons). Changing $\phis$ amounts to simulate the behaviour of identical particles of different nature stemming from sLOCC: therefore, the ability of our setup to directly measure $\phis$ gives the ability to directly detect the EP of any type of particles.
Renaming $\phi$ the simulated exchange phase, we get
    \begin{equation}
        \langle\hat{O}\rangle
        \equiv\bra{\psi_f}\hat{O}\ket{\psi_f}
        =\sin(2\beta)\cos\phi,
    \end{equation}
    from which $\phi$ can be easily extrapolated.
	
We set $\beta=45^\circ$ and $\phi=0$ to have two maximally indistinguishable photons (bosons), generating maximum entanglement $\ket{\psi_\text{LR}}=(\ket{{\rm L}H,{\rm R}V}+\ket{{\rm L}V,{\rm R}H})/\sqrt{2}$ with a fidelity of $0.99\pm0.01$.
Unavoidable experimental errors prevent the achievement of maximum indistinguishability and result in a non-optimal performance of the real setup.
Following the method used in Ref.\tilde\cite{tschernig2021direct}, we treat such errors as a constant factor affecting the final experimental results. We assume that the experimentally prepared state is the desired (ideal) one with probability $F$, while errors give rise to a spoiled state with probability $1-F$. Within this model, the spoiled state does not contribute to the expectation value of $\hat{O}$, leading to the experimentally measured expectation value $\langle \hat{O}\rangle_e=F\langle\hat{O}\rangle_i$.
By preparing several states of Eq.\tilde\eqref{exppsilr} for different values of $\phi$, we use quantum tomography to estimate the error probability to be $F=0.977$ (see Appendix \ref{app:III}).
	
Fig.~\ref{fig3}\textbf{b} and \ref{fig3}\textbf{c} show the coincidence counts $\hat{n}_{13}$, $\hat{n}_{14}$, $\hat{n}_{23}$ and $\hat{n}_{24}$ detected for (real) bosons and (simulated) fermions, respectively. Treating experimental errors as explained above, we obtain $\phi_b=0.04\pm0.06$ for bosons and $\phi_f=3.12\pm0.05$ for fermions, confirming their expected EPs. Here, the standard deviation is estimated from experimental data via Monte Carlo methods.

By varying the HWP angle $\beta$, we measure several simulated phases (including anyonic ones) for various spatial overlaps to provide deeper insights on the role played by spatial indistinguishability in our scheme.
Results are shown in Fig.~\ref{fig3}\textbf{d}. Measured values of $\langle\hat{O}\rangle$ are given as a function of $\cos\phi$, with $\phi$ obtained via tomographic measurements, for different degrees of spatial overlap and, hence, of spatial indistinguishability $\mathcal{I}= -\sin^2\beta \log_2 (\sin^2\beta)-\cos^2\beta \log_2 (\cos^2\beta)$~\cite{indistentanglprotection}.
In particular, $\beta=45^\circ$ corresponds to maximum spatial overlap ($\mathcal{I}=1$), while $\beta=30^\circ$, $\beta=20^\circ$, $\beta=10^\circ$ are associated to partial spatial overlaps ($\mathcal{I}<1$).
Experimental results are reported with different markers, solid lines represent the ideal theoretically expected values $\langle \hat{O}\rangle_i$, while dashed lines correspond to the  values $\langle\hat{O}\rangle_e$ considering experimental errors.
Notice that when $\mathcal{I}$ decreases, the ranges of values of $\langle\hat{O}\rangle$ decrease accordingly, leading to a lower sensibility. Spatial indistinguishability acts as a sensibility regulator ruling the range of measured values.

	\begin{figure}[!t]
		\centering
		\includegraphics[width=0.48\textwidth]{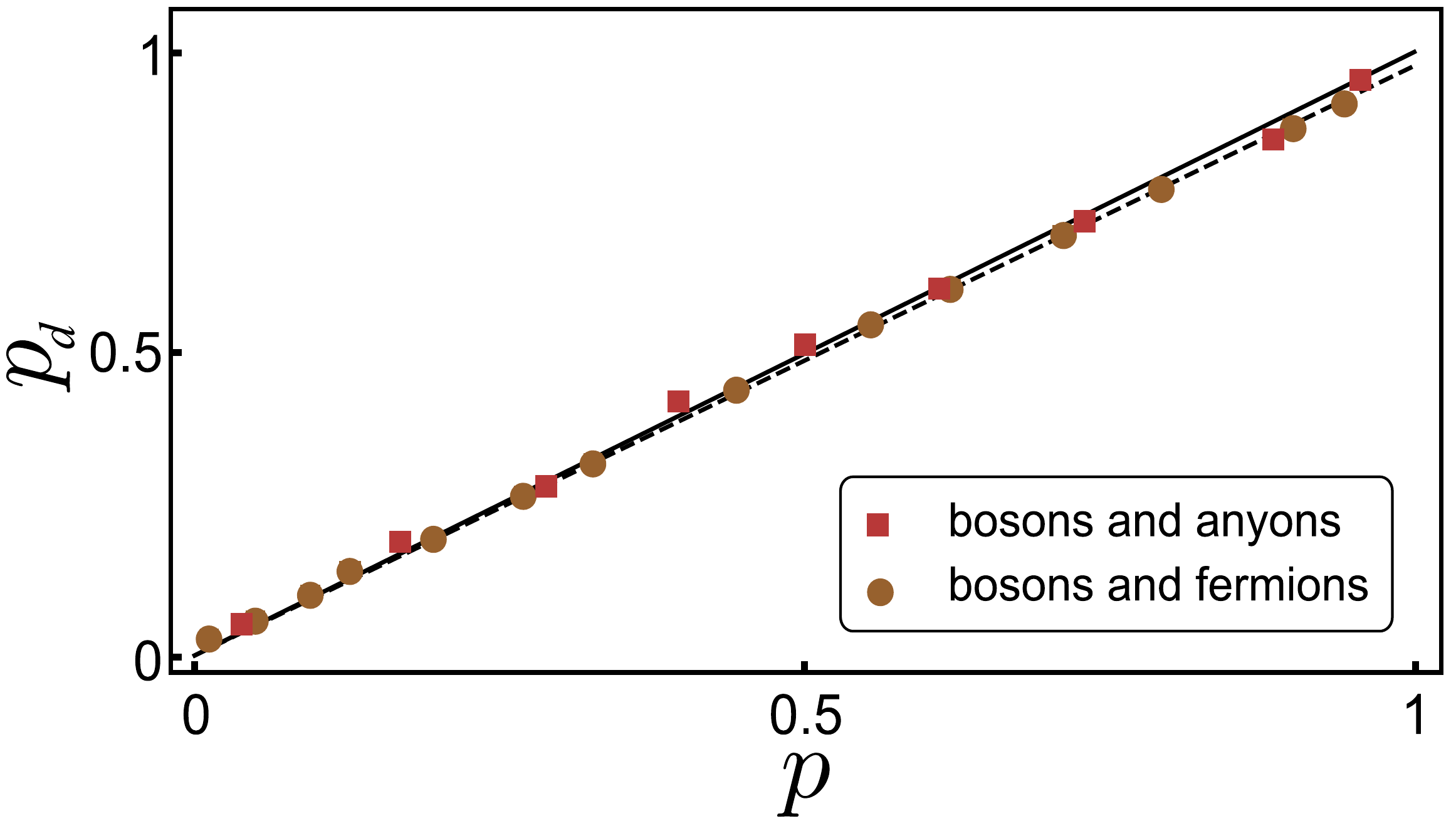}
		\caption{Probability distribution $p_d$ for a mixture $\rho$ of two types of particles measured by our procedure versus the simulated value $p$. Experimental results for a mixture of bosons and anyons with EP $\phi=\pi/2$ (red markers), and for a mixture of bosons and fermions (brown markers). Solid lines represent the theoretically expected ideal values, dashed lines the theoretical values when noise is considered. Errorbars are too small to be visible.}
		\label{fig6}
	\end{figure}
	
As an extension of our framework, we analyze the scenario where the input is a flux of particle pairs whose exchange phase is known to be either $\phi_1$ with probability $p$ or $\phi_2$ with probability $1-p$. Each two-particle state is thus given by the classical mixture
\begin{equation}
\label{mixture}
    \rho=p\ket{\psi_1}\bra{\psi_1}+(1-p)\ket{\psi_2}\bra{\psi_2},
\end{equation}
where $\ket{\psi_1}=\cos{\beta}\ket{{\rm L} H,{\rm R} V}+e^{i\phi_1}\sin{\beta}\ket{{\rm L} V,{\rm R} H}$ and $\ket{\psi_2}=\cos{\beta}\ket{{\rm L} H,{\rm R} V}+e^{i\phi_2}\sin{\beta}\ket{{\rm L} V,{\rm R} H}$. We now want to exploit our procedure to estimate the probability distribution $p$ of the two types of particles by directly measuring their EPs.
To prepare $\rho$, we replace the dotted box in Fig.~\ref{expe}\textbf{a} with the unbalanced interferometer shown in Fig.~\ref{expe}\textbf{c}. Here, a couple of HWPs is followed by two BDs which vertically split each beam in two. The two upper arms (orange lines) are used to simulate the particles with EP $\phi_1$, while the two lower arms (blue lines) are used to simulate the particles with EP $\phi_2$. By changing the angles of the two HWPs before the BDs, different probability distributions $p$ are simulated, while the EPs $\phi_1$ and $\phi_2$ are controlled using a pair of PAs.
One HWP set at $0^\circ$ or $45^\circ$ is inserted on the four paths to adjust the polarizations. Finally, another BD regroups the two upper arms and the two lower arms into two beams which meet at a beam splitter (BS); one output of the BS is blocked, while the other generates the desired classical mixed state of Eq.\tilde\eqref{mixture}.
The expectation value 
		$\langle\hat{O}\rangle={\rm Tr}[\rho\,\hat{O}]
		=p\bra{\psi_1}\hat{O}\ket{\psi_1}+(1-p)\bra{\psi_2}\hat{O}\ket{\psi_2}$
is measured as before. For simplicity, we assume that the values of $\phi_1$ and $\phi_2$ are given as prior information, so as to compute the expectation values $\bra{\psi_j}\hat{O}\ket{\psi_j}$ ($j=1,2$) and reduce $\langle\hat{O}\rangle$ to one linear equation in $p$. Notice that, if this is not the case, the values of $\phi_1$ and $\phi_2$ can nonetheless be obtained by our procedure to directly measure them on a sufficiently big sample of particle pairs. We fix $\beta=45^\circ$. We start with $\phi_1=0$ and $\phi_2=\pi/2$ to simulate a classical mixture of bosons and anyons. We create different probability distributions by rotating the two input HWPs as shown in Fig.~\ref{expe}\textbf{c}. Then, we set $\phi_2=\pi$ to simulate a mixture of bosons and fermions and repeat the experiment. The results are reported in Fig.\tilde\ref{fig6}, where the observed probability distributions $p_d$, inferred by the measured values of $\langle\hat{O}\rangle$, are shown versus the simulated $p$. Excellent agreement with theoretical predictions is observed. This latest experiment demonstrates how our procedure can be used to obtain information on the probability distribution $p$ for a mixture of two known different types of particles. If the number of types of particles is increased or unknown, a complete characterization of the incoming flux can still be given by directly measuring the various EPs of the particle pairs composing a large enough sample.


As an outlook, it would be interesting to apply our setup in non-optical platforms to achieve the first direct measurement of real (not simulated) fermionic and anyonic EPs. In fact, our scheme can be translated to any platform implementing linear optics, such as electronic optics\tilde\cite{bauerle2018coherent}. Also, quantum dots appear promising for on demand-generation of single electrons\tilde\cite{feve2007demand}, including their initialization and coherent control\tilde\cite{press2008complete}, whose indistinguishability degree can be adjusted by quantum point contacts acting as electronic beam splitters\tilde\cite{bocquillon2013coherence}.
	
In summary, we have experimentally shown that the sLOCC framework is inherently amenable for direct measurement of the EP of indistinguishable particles. 
Particle statistics in the measured state is entirely due to the spatial indistinguishability achieved via the deformation of particle wave packets. The sLOCC process functions as a trigger making the EP directly accessible within the generated entanglement. For this reason, physical exchange of particles and related geometric phase do not occur here, in contrast with the technique previously adopted~\cite{tschernig2021direct} to measure the bosonic EP of photons.	
Our procedure works for bosons, fermions and anyons. We have judiciously designed the optical setup to simulate various particle statistics: differently from other methods used for this aim in photonic quantum walks\tilde\cite{matthews2013observing,sansoni2012two}, we have manually injected different EPs by accurately tuning a phase adjustment, always observing agreement between measured values and predictions. 
Our apparatus has confirmed the real bosonic (symmetric) nature of photons, enclosing the result of Ref.~\cite{tschernig2021direct}.  
We have also proven that repeated measurements of the EP permit to reconstruct the probability distribution for statistical mixtures of states of particles of different natures. 
Our work provides a general scheme to directly explore the symmetrization principle and the role of particle statistics in various contexts.

\section*{Acknowledgements} 
This work was supported by the National Key Research and Development Program of China (Grant NO. 2017YFA0304100), National Natural Science Foundation of China (Grant NOs. 11821404, 11774335, 61725504, 61805227, 61975195, U19A2075), Anhui Initiative in Quantum Information Technologies (Grant NOs.\ AHY060300 and AHY020100), Key Research Program of Frontier Science, CAS (Grant NO.\ QYZDYSSW-SLH003), Science Foundation of the CAS (NO. ZDRW-XH-2019-1), the Fundamental Research Funds for the Central Universities (Grant NO. WK2030380017 and WK2470000026). R.M. thanks support from NSERC, MEI and the CRC program in Canada. R.L.F. acknowledges support from ``Sistema di Incentivazione, Sostegno e Premialit\`a della Ricerca Dipartimentale'' of the Department of Engineering, University of Palermo. M.P. and R.L.F. would like to thank Farzam Nosrati for insightful discussions.

	\appendix
	
	\section{Theoretical framework}\label{app:Theory}

	The conceptual procedure is depicted in Fig.\tilde\ref{expe}(a).
Let us take a pair of two-level identical particles independently prepared and initially uncorrelated, whose spatial wave functions and pseudospins are respectively $\varphi,\,\uparrow$ and $\varphi^\prime,\,\downarrow$. In the no-label formalism, we write this state as $\ket{\psi_\textrm{in}}=\ket{\varphi\uparrow,\varphi^\prime\downarrow}$. Then, a deformation operation $\ket{\varphi}\rightarrow\ket{\varphi_\textrm{D}}$, $\ket{\varphi^\prime}\rightarrow\ket{\varphi_\textrm{D}^\prime}$ is performed\tilde\cite{indistentanglprotection,Piccolini_2021} to distribute the spatial wave functions over two distinct regions L and R in a controllable way, thus transforming $\ket{\psi_\textrm{in}}$ into $\ket{\psi_\textrm{D}}=\ket{\varphi_\textrm{D}\uparrow,\varphi^\prime_\textrm{D}\downarrow}$, where
    \begin{equation}
  \ket{\varphi_\textrm{D}}=l\ket{\rm L}+r\ket{\rm R},
            \quad
            \ket{\varphi_\textrm{D}^\prime}=l^\prime\ket{\rm L}+r^\prime\ket{\rm R}.
    \end{equation}
Here, the coefficients $l=\langle \rm L|\varphi_D\rangle$, $l^\prime=\langle \rm L|\varphi_D^\prime\rangle$, $r=\langle\rm R|\varphi_D\rangle$ and $r^\prime=\langle\rm R|\varphi_D^\prime\rangle$ are the tunable probability amplitudes of finding the particle whose spatial wave function is $\varphi_D$ or $\varphi_D^\prime$ in the site L and R, respectively.

To implement the sLOCC measurement we perform the post-selected detection of the states where exactly one qubit per region is recorded. In total, this last step amounts to projecting the state $\ket{\psi_D}$ onto the two particle basis
	$\mathcal{B}_{\rm LR}
	=\{\ket{\rm L\uparrow,\rm R\uparrow},\ket{\rm L\uparrow,\rm R\downarrow},\ket{\rm L\downarrow,\rm R\uparrow},\ket{\rm L\downarrow,\rm R\downarrow}\}$
	via the projection operator $\hat{\Pi}_{\rm LR}=\sum_{\sigma,\tau=\uparrow,\downarrow}\ket{\rm L\sigma,\rm R\tau}\bra{\rm L\sigma,\rm R\tau}$.
	
%

We recall that the two particles in the state $\ket{\psi_D}$ are indistinguishable to the eyes of the detectors. This means that it is not possible to know the region of space where each detected constituent is coming from. Such no which-way information is encoded in the result of the sLOCC operation, which is easily computed to be the (normalized) two-particle entangled state
	\begin{equation}
		\label{psilr}
		\begin{aligned}
			\ket{\psi_{\rm LR}}
			&=\frac{\hat{\Pi}_{\rm LR}\ket{\psi_D}}
			{\sqrt{\bra{\psi_D}\hat{\Pi}_{\rm LR}\ket{\psi_D}}}\\
			&=\frac{lr^\prime\ket{\rm L\uparrow,\rm R\downarrow}+e^{i\phi}rl^\prime\ket{\rm L\downarrow,\rm R\uparrow}}
			{\sqrt{|lr^\prime|^2+|rl^\prime|^2}},
		\end{aligned}		
	\end{equation}
generated with probability $P_{\rm LR}=|lr^\prime|^2+|rl^\prime|^2$ \cite{slocc}. The naturally emerged phase $\phi$ in Eq.\tilde\eqref{psilr} is exactly the relative EP we want to measure (Fig.\tilde\ref{expe}(a)). In fact, it is fundamentally contained in the probability amplitudes $\bra{\chi_{\mathrm{L}},\chi_{\textrm{R}}}\psi_\mathrm{D}\rangle=\bra{\chi_{\mathrm{L}}}\chi_\textrm{D}\rangle \bra{\chi_{\mathrm{L}}}\chi'_\textrm{D}\rangle +\eta \bra{\chi_{\mathrm{L}}}\chi'_\textrm{D}\rangle \bra{\chi_{\mathrm{R}}}\chi_\textrm{D}\rangle$ \cite{nolabelappr}, where $\chi_{\mathrm{L}}=\mathrm{L}\sigma$, $\chi_{\mathrm{R}}=\mathrm{R}\tau$, $\chi_\textrm{D}=\varphi_\textrm{D}\uparrow$, $\chi'_\textrm{D}=\varphi'_\textrm{D}\downarrow$ and $\eta=e^{i\phi}$ is the particle statistics parameter. It is worth to highlight that the state $\ket{\psi_{\rm LR}}$, resulting from the sLOCC process, describes two particles occupying two distinct regions of space, thus being now distinguishable and individually addressable.  The spatial indistinguishability $\mathcal{I}$ under sLOCC associated to the state $\ket{\psi_\mathrm{D}}$, and thus to the state $\ket{\psi_\mathrm{LR}}$, is given by \cite{indistentanglprotection}
\begin{eqnarray}
\mathcal{I}&=&-\frac{|lr'|^2}{|lr'|^2+|l'r|^2}\log_2\frac{|lr'|^2}{|lr'|^2+|l'r|^2}\nonumber\\
&&-\frac{|l'r|^2}{|lr'|^2+|l'r|^2}\log_2\frac{|l'r|^2}{|lr'|^2+|l'r|^2}.
\end{eqnarray}

		\begin{figure*}[t!]
	\centering
	\includegraphics[width=\textwidth]{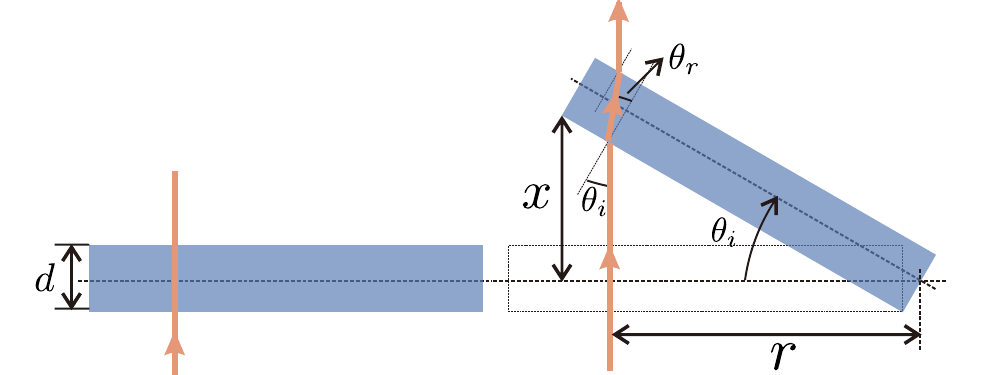}
	\caption{\textbf{Sketch of the tilted experimental setup.}}
	\label{fig:tilted}
\end{figure*}
	
In general, the state of Eq.~\eqref{psilr} represents a quantum superposition of two-particle states whose relative phase contains the EP of the particles. Notice that one of the major difficulties in directly measuring the particle statistical phase consists in creating quantum interference between a given state and its counterpart where particles have been physically exchanged \cite{lo2021directly}. A so-called state-dependent transport protocol has been first engineered to this aim \cite{roos2017} and successively realized with photons \cite{tschernig2021direct}. On the other hand, in our scheme, the fundamental EP straightforwardly appears as a natural consequence of spatial overlap at separated regions plus the sLOCC procedure, being hence amenable to be directly measured via individual operations on the particles. We then proceed by rotating the pseudospin of both qubits by $\pi/4$. Given the single particle operator
	\begin{equation}
	\label{rotation}
		\hat{M}_{X}
		=\frac{1}{\sqrt{2}}
		\left(\begin{array}{cc}
			1 & -1 \\
			1 & 1
		\end{array}\right),
	\end{equation}
performing such operation on the region $X=L$, $R$, the resulting state is given by $\ket{\psi_f}=\hat{M}_{\rm L}\otimes\hat{M}_{\rm R}\ket{\psi_{\rm LR}}$.
Finally, a direct measurement of the pseudospin along the z-axis in both regions $L$ and $R$ provides information about the EP. Indeed,  we find that
	\begin{equation}
		\label{theory}
		\langle\psi_{f}\left|\hat{\sigma}_{\rm L}^{z} \otimes \hat{\sigma}_{\rm R}^{z}\right| \psi_{f}\rangle=\frac{2lr^\prime rl^\prime}{|lr^\prime|^2+|rl^\prime|^2}\cos\phi,
	\end{equation}
where we have taken the coefficients $l, r, l^\prime, r^\prime$ to be real since we are able to directly control the distribution of the initial spatial wave functions over L and R during the preparation of the state $\ket{\psi_\textrm{D}}$.
By knowing such amplitudes, it is thus possible to recover the value of the EP from repeated pseudospin measurements along the z-axis.
	
Remarkably,  the role of spatial indistinguishability $\mathcal{I}$ clearly emerges from Eq.~\eqref{theory}: indeed, as its value varies from $\mathcal{I}=1$ (maximum indistinguishability, obtained, e.g., when $l=r=l'=r'=1/\sqrt{2}$) to $\mathcal{I}=0$ (distinguishable particles, e.g., when $l=r'=1,\,l'=r=0$) \cite{indistentanglprotection},  the values assumed by Eq.~\eqref{theory} continuously change from $\cos\phi$ to zero,  correspondingly. It follows that spatial indistinguishability is not only an essential element for measuring the EP with our procedure, but it also acts as a sensibility regulator which tunes our ability to access the value of $\phi$.

	\section{Introduction of phase adjustment}\label{app:PA}

	For a better and intuitive understanding of subtle adjustment of EP $\phi$, we would like to build the intuitive geometrical relationship between the distance $x$ ($mm$) of the moving plate and corresponding EP $\phi$. This would help us to further connect the moving distance $x$ with the direct observable $\hat{O}$,  by exploiting our experimental setup's capability to obtain the exact phase of EP directly.
	
	Here the corresponding EP of simulated identical particles is ranging from $0$ to $\pi$. The necessary experimental initialization starts from the adjustment of $\phi=0$ with two photons separately passing perpendicularly two thin plates of fused quartz (both having the same thickness $d$), of which one is motionless and the other can be rotated by a small angle. Such a rotation is driven by a movable plate (MP). The thickness of $d$ is about $200\ \mu m$, so that the moving part has no influence on the parts of the setup which follow. At each displacement, the tomography procedure is performed so as to construct the corresponding density matrix and, furthermore, confirm the relative phase $\phi$ \cite{daniel_2001}. The experimental results are represented with the black dots in Fig.~3{\bf a} of the main text (lower panel), where the associated errorbars are too small to be seen.

	
Here, we determine the theoretical predictions based on our experimental setup. As shown in Fig.~\ref{fig:tilted}, the incident angle $\theta_i$ and refractive angle $\theta_r$ satisfy the refraction law where the index of glass sheet is $n=1.5$ and the index of air is $n_0=1$. We can assume $x=r\sin\theta_i$ for a small angle $\theta_i$.
The relationship between the moving distance $x$ and the phase $\phi$ is
	 \begin{eqnarray}
	 \label{xd}
	 	      \phi &=&\frac{2\pi}{\lambda}nd\left(\frac{1}{\sqrt{1-(\frac{\sin\theta_i n_0}{n})^2}}-1\right)\nonumber\\
	 	      &=&\frac{2\pi}{\lambda}nd\left(\frac{1}{\sqrt{1-(\frac{x}{r n})^2}}-1\right),
	 \end{eqnarray}
	 where $\lambda$ corresponds to the wavelength of the photon.  Moreover, we find that the parameters $d$ and $r$ are $d=199.94\pm1.43\ \mu m$ and  $r=102.36\pm0.91\ mm$. These numbers are in agreement with the measured values, and fit well with our experimental results presented in Fig. 3{\bf a} of the main text. This means we can straightforwardly obtain the EP $\phi$ through the displacement of MP $x$ $(mm)$, which is a more intuitive quantity than $\phi$. Also, we can exploit the direct measurement results between observable $\hat{O}$ and $x$ instead of $\hat{O}$ and $\phi$, see Fig.~4{\bf a} in the main text. Based on Eq. \ref{xd}, $\phi$ could be adjusted to be larger than $\pi$; however, considering its periodicity, it can be transformed to a value which is within the range [0,$\pi$].
	
	

	\section{Treatment of the experimental errors and prediction of the setup performances} \label{app:III}
	
	The temporal indistinguishability characterizing the two photons in our setup is evaluated by measuring the Hong-Ou-Mandel interference dip, showing a visibility of $97.7\%$. Such uncomplete indistinguishability is the result of the unavoidable environmental decoherence and of the somehow limited performances of our experimental setup due to, e.g.,  white noise effects and dark counts, leading to the generation of states which slightly deviate from the ideal ones. Similarly to Ref.~\cite{tschernig2021direct}, we model these experimental errors as a constant factor and compute the estimated performance of our setup.
	
	The ideal state we would like to prepare is
	\begin{equation}
	   \ket{\psi_{\rm LR}}=\cos{\beta}\ket{{\rm L} H,{\rm R} V}+e^{i\phi}\sin{\beta}\ket{{\rm L} V,{\rm R} H},
	   \label{1}
	\end{equation}
	where $\phi$ represents the simulated EP while $\beta$ characterizes the degree of spatial indistinguishability. Denoting with $\rho_i$ the corresponding pure state density matrix, i.e. $\rho_i=\ket{\psi_{\rm LR}}\bra{\psi_{\rm LR}}$, we model the experimental errors as if they would lead to the generation of the ideal state $\rho_i$ with probability $F$. Naming $\rho_n$ the otherwise achieved noisy state obtained with probability $1-F$, the setup thus generates the mixed state
	\begin{equation}
	\label{rhoe}
	    \rho_e=F\rho_i+(1-F)\rho_n.
	\end{equation}
    We consider the noisy part as composed of two contributions: a state $\rho_{n1}$ accounting for the white noise due to the accidental errors
	\begin{eqnarray}
	 \rho_{n1}&=&\frac{1}{4}(\ket{{\rm L} H,{\rm R} H}\bra{{\rm L} H,{\rm R} H}+
	 \ket{{\rm L} H,{\rm R} V}\bra{{\rm L} H,{\rm R} V}\nonumber\\
	 &&+\ket{{\rm L} V,{\rm R} H}\bra{{\rm L} V,{\rm R} H}+\ket{{\rm L} V,{\rm R} V}\bra{{\rm L} V,{\rm R} V}),
	\end{eqnarray}
	and a state $\rho_{n2}$ accounting for decoherence effects
	\begin{equation}
	    \rho_{n2}=\frac{1}{2}(\ket{{\rm L} H,{\rm R} V}\bra{{\rm L} H,{\rm R} V}+\ket{{\rm L} V,{\rm R} H}\bra{{\rm L} V,{\rm R} H}).
	\end{equation}
	The complete noisy state generated is thus given by
		\begin{equation*}
	    \rho_n=a\rho_{n1}+b\rho_{n2},
	\end{equation*}
	where the coefficients $a$ and $b$ are such that $a+b=1$.
	
	It is now easy to show that, once rotated as described in the main text, the noisy component does not contribute to the expectation value of the observable $\hat{O}=\hat{\sigma}_{\rm L}^{z} \otimes \hat{\sigma}_{\rm R}^{z}$ we want to measure. As a consequence, the only relevant effect of the experimental errors within this model is to reduce the visibility of the two indistinguishable photons, meaning that the experimental results $\langle \hat{O}\rangle_e$ are related to the ideal ones $\langle \hat{O}\rangle_i$ by $\langle\hat{O}\rangle_e=F*\langle\hat{O}\rangle_i$.
	 
    We now want to estimate the parameter $F$.
	To do so, we use quantum tomography to experimentally reconstruct $\rho_e$ for different states generated while varying $\phi$ from $0$ to $\pi$ \cite{daniel_2001}.
	This allows us to compute $\phi$ and $\beta$, from which the ideal state \eqref{1} can be reconstructed. By preparing several experimental states and obtaining the corresponding groups of $\phi$ and $\beta$ we use \eqref{rhoe} to get $F=0.977$.
	As for the noisy part $\rho_n$, composed of $\rho_{n1}$ and $\rho_{n2}$, the values of $a$ and $b$ have a little difference. 
	Remarkably, we find that the parameter $F$ is affected by fluctuations whose magnitude is of the order of $10^{-3}$.
	
	The above presented treatment of the experimental errors is exploited in the main text to perform the direct measurement of the EP with an higher accuracy (see main text).




\section{Further experimental plots}\label{app:IV}

\begin{figure}[t!]
	\centering
	\includegraphics[width=0.48\textwidth]{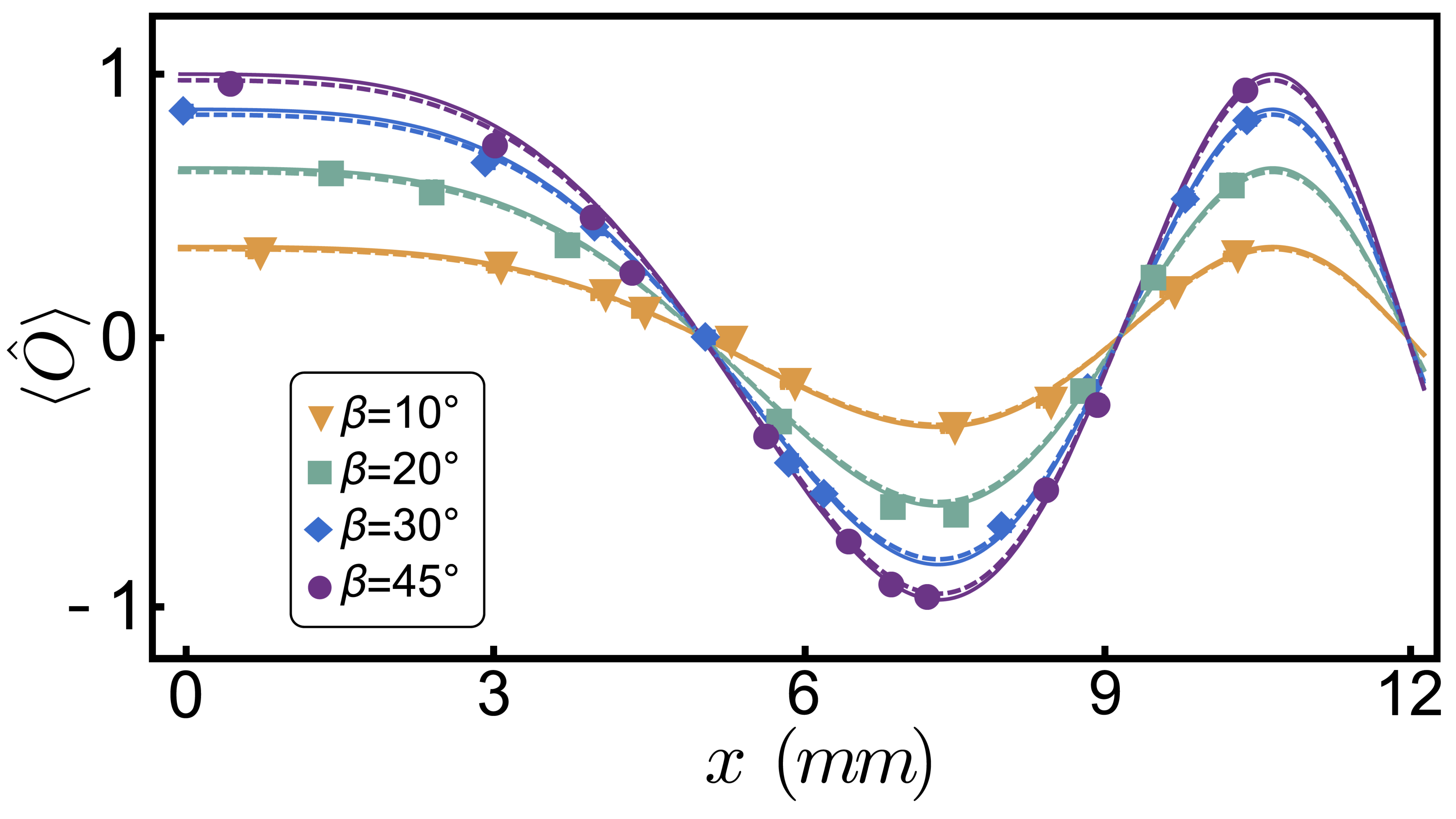}
		\caption{\textbf{Experimental results for different simulated types of particles.} \textbf{a.} Measured values of $\langle\hat{O}\rangle$
		versus the displacement $x$ (mm) of the moving plate generating a relative phase $\phi$ in the range from $0$ to $\pi$.  
		Results are reported for different values of $\beta$ (different colors), corresponding to different degrees of spatial indistinguishability. The solid lines represent the theoretically expected results in the ideal (no noise) scenario, while the dashed lines show the theoretical values when noise is taken into consideration. Experimentally measured values are represented with markers.}
		\label{fig41}
\end{figure}

\begin{figure}[t!]
		\centering
		\includegraphics[width=0.48\textwidth]{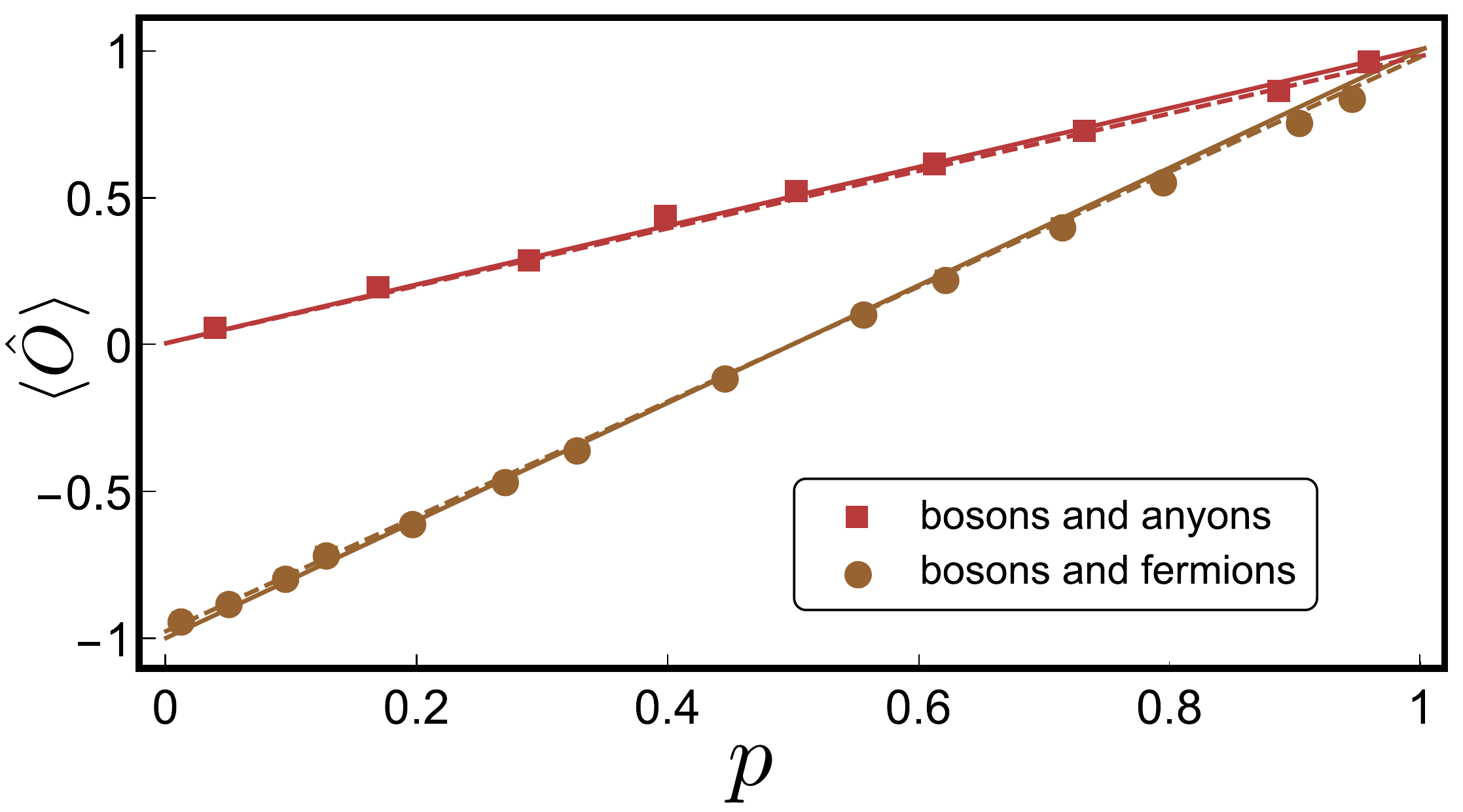}
		\caption{Relation between $\langle \hat{O}\rangle$ and the simulated probability distribution $p$ of a mixture of bosons and anyons with EP $\phi=\pi/2$ (red) and one of bosons and fermions (brown). Solid lines represent the theoretically expected ideal values, dashed lines represent the theoretically expected values when noise is considered, while markers represent the experimentally obtained results. Errorbars are too small to be visible.}
		\label{fig5}
	\end{figure}	

By changing the angle $\beta$ of the HWP, we measure various simulated phases (including anyonic ones) for different values of spatial overlap in order to give further insights on the role played by spatial indistinguishability in our framework.
Results are shown in Fig.~\ref{fig41}. Here, the measured values of $\langle\hat{O}\rangle$ are given directly as a function of the movable plate's displacement for different degrees of spatial overlap and, thus, of spatial indistinguishability $\mathcal{I}$. The latter is defined by an entropic expression in terms of the probabilities of finding each particle in a given region~\cite{indistentanglprotection} and, in our experiment, reads $\mathcal{I}= -\sin^2\beta \log_2 (\sin^2\beta)-\cos^2\beta \log_2 (\cos^2\beta)$.
We recall that $\beta=45^\circ$ corresponds to a maximum spatial overlap (i.e., maximum spatial indistinguishability $\mathcal{I}=1$), while $\beta=30^\circ$, $\beta=20^\circ$, $\beta=10^\circ$ denote partial spatial overlaps ($\mathcal{I}<1$).
The experimental results are reported with markers of different types, while the solid lines represent the ideal theoretically expected values $\langle \hat{O}\rangle_i$.
Finally, the dashed lines correspond to the values $\langle\hat{O}\rangle_e$ expected when taking in consideration the experimental errors.
Notice that, as previously discussed, when the degree of spatial indistinguishability decreases the ranges of values assumed by $\langle\hat{O}\rangle$ decrease accordingly, thus leading to a lower sensibility.

In performing the second experiment for the classical mixture of two types of particles, we set $\beta=45^\circ$ for simplicity, as reported in the main text. To begin with, we set $\phi_1=0$ and $\phi_2=\pi/2$ to simulate a classical mixture of bosons and anyons. We simulate different probability distributions by rotating the two input HWPs shown in
Fig. 1 of the main text. The expectation value of $\hat{O}$, given by
\begin{equation}
\label{expO}
		\langle\hat{O}\rangle={\rm Tr}[\rho\,\hat{O}]
		=p\bra{\psi_1}\hat{O}\ket{\psi_1}+(1-p)\bra{\psi_2}\hat{O}\ket{\psi_2},
\end{equation}
is measured as in the pure states scenario. For simplicity, we assume that the values of $\phi_1$ and $\phi_2$ are given as prior information, allowing us to compute the expectation values $\bra{\psi_j}\hat{O}\ket{\psi_j},\,j=1,2$ and reducing Eq.\tilde\eqref{expO} to one linear equation in $p$. Notice that, if this is not the case, the values of $\phi_1$ and $\phi_2$ can nonetheless be obtained by exploiting our procedure to directly measure them on a sufficiently big sample of particle pairs.

The obtained results are represented with red markers in Fig.\tilde\ref{fig5}, where $\langle\hat{O}\rangle$ is plotted against the simulated probability $p$. Here, the efficacy of our setup is appearent when compared to the red solid line representing the theoretically expected values of $\langle\hat{O}\rangle$ computed using Eq.\tilde\eqref{expO}. As before, the dashed line represents the theoretical values expected when considering the action of noise, i.e. $\langle\hat{O}\rangle_e=F\,\langle\hat{O}\rangle$.
Then, we set $\phi_2=\pi$ to simulate a mixture of bosons and fermions and repeat the experiment. The results, displayed in Figs.\tilde\ref{fig5} with brown markers and lines, are once again in good agreement with our theoretical predictions.


%

\end{document}